\documentstyle[epsfig,twoside,fleqn,espcrc2]{article}

\newcommand{\AmS}{{\protect\the\textfont2
  A\kern-.1667em\lower.5ex\hbox{M}\kern-.125emS}}

\title{
\vspace{-8mm}
\rightline{\small HUB--EP--97/59}
\vspace{-2mm}
\rightline{\small September 9, 1997}
Hot electroweak matter near to the endpoint of the phase transition
}
\author{
  M.~G\"urtler\address{Institut f\"ur Theoretische Physik, Universit\"at
    Leipzig, Augustusplatz 10-11, D-04109 Leipzig, Germany},
  E.-M.~Ilgenfritz\address{Institut f\"ur Physik, Humboldt-Universit\"at zu
    Berlin, Invalidenstr. 110, D-10115 Berlin, Germany},
  A.~Schiller$\mathrm{^a}$\thanks{Talk presented by A.~Schiller},
  and C.~Strecha$\mathrm{^a}$}
\begin{document}
\begin{abstract}
  The electroweak phase transition is investigated near to its endpoint in the
  framework of an effective three--dimensional model.  We measure the very
  weak interface tension with the tunneling correlation length method. First
  results for the mass spectrum and the corresponding wave functions in the
  symmetric phase are presented.
\end{abstract}

\maketitle

\section{The model}

The model under consideration is the $3d$ $SU(2)$--Higgs model with the
lattice action
\begin{eqnarray}
  S & =& \beta_G \sum_p \big(1 - {1 \over 2}{\mathrm tr} U_p \big)  
\nonumber \\
  &&
- \beta_H \sum_{x,\mu}
  {1\over 2}{\mathrm tr} (\Phi_x^+ U_{x, \mu} \Phi_{x + \mu})
  \nonumber \\
  &  & +  \sum_x  \big( \rho_x^2 + \beta_R (\rho_x^2-1)^2 \big)  \ \ ,
  \label{eq:latt_action}
\end{eqnarray} 
$\rho_x^2= (1/2) {\mathrm tr} (\Phi_x^+ \Phi_x$).  Our conventions and update
details can be found in \cite{wirNP97}.  Relations between the parameters of
this effective model and the physical quantities Higgs mass and temperature
are derived in \cite{generic}.

\section{Interface tension}
\label{sec:interface}

The interface tension $\alpha$ characterises the strength of a first order
phase transition. We apply the tunnelling correlation method \cite{muenster}
to determine its value for a Higgs mass of about $70$~GeV by studying the
inverse correlation length $m_{\mathrm{lat}}=m_{\mathrm {gap}}a $ (given by
the correlation function of the Higgs operator $\rho_x^2$
along $z$) in an elongated lattice $L^2 \times
L_z$ as function of the transverse area $a^2L^2$.  The scaling of
$m_{\mathrm{gap}}$ at the critical point is predicted for a $\phi^4$-model as
$ m_{\mathrm{gap}} a L \propto x \exp\left(-x \right)$ with $x= \alpha a^2 L^2
/T_c$ ~\cite{muenster} by considering quadratic fluctuations around the kink
solution in a semiclassical expansion.
\begin{figure}[!htb]
  \centering
  \epsfig{file=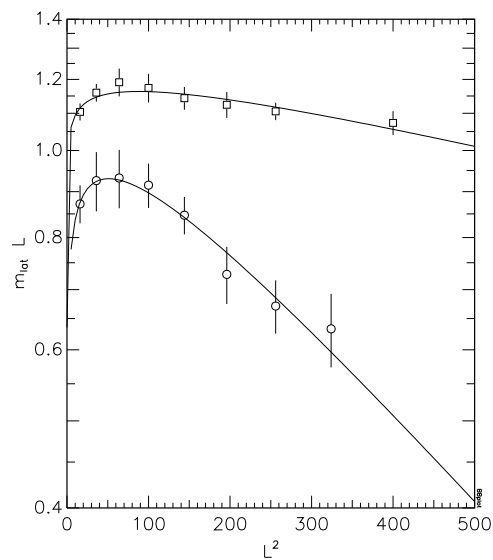,width=6.5cm,angle=0}
  \vspace{-1cm}
  \caption{\sl Fits for the inverse tunneling correlation lengths at 
    $M_H^*=57$~{\rm GeV} (circles) and $M_H^*=70$~{\rm GeV} (squares)}
  \label{fig:three}
  \vspace{-9mm} 
\end{figure}

The power of $L$ in the scaling law for $m_{\mathrm{gap}}$ is eventually
modified in higher orders.  More general fits according to
\begin{eqnarray}
  \label{eq:mod_scaling}
    m_{\mathrm{gap}} a L =c L^\gamma x \exp\left(-x \right)
\end{eqnarray}
give the $3d$ interface tension $\alpha_3/g_3^4=0.0049(18)$ (using
$x=(\alpha_3/g_3^4) \, L^2 (4/\beta_G)^2$ for $3d$ units) at $M_H^*=70$~GeV.

As a check for the applicability of this method we have repeated the
simulation for $M_H^*=57$~GeV and obtained $\alpha_3/g_3^4=0.0224(56)$ in
agreement with \cite{kajantie}.  The measurements are shown in
Fig.~\ref{fig:three}.
\begin{figure}[!htb]
  \centering
  \epsfig{file=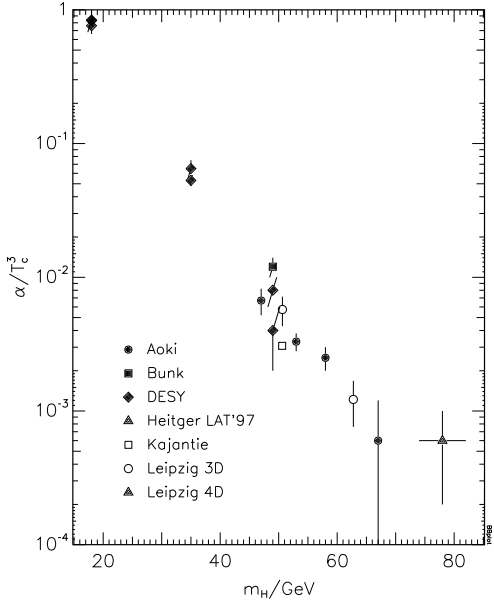,width=6.5cm,angle=0}
  \vspace{-11mm}
  \caption{\sl Interface tension for several Higgs masses; for
    references see {\rm[5,6]}.}
  \label{fig:surf}
  \vspace{-9mm}
\end{figure}

Fig.~\ref{fig:surf} collects published results for the interface tension
(scaled by critical temperature) $\alpha/T_c^3$ of the $SU(2)$--Higgs model.
In order to make this useful we had to recalculate the $3d$ results to $4d$
conventions (adjusting to a renormalised gauge coupling 0.58 characteristic
for the $4d$ lattice data).  This comparison further supports the validity of
dimensional reduction.

\section{Wavefunctions and excited states}

Knowledge of the mass spectrum of screening states in the symmetric phase is
important to describe the high $T$ phase by non-perturbative analytic models.
The wave functions may characterise the change between the phases.  First
studies have been reported recently \cite{Philipsen}.

To study ground state {\sl and} excited states (and possibly wave functions)
one has to use cross-correlations between operators ${\cal {O}}_{i}$ from a
complete set in a given $J^{PC}$ channel.  In the transfer matrix formalism
they describe the spectral decomposition ($\Psi_i^{(n)}=\langle \mathrm{vac}|
{\cal{O}}_i | {\bf \Psi}^{(n)}\rangle$, $|{\bf \Psi}^{(n)}\rangle$ being the
zero momentum energy eigenstates)
\begin{eqnarray}
 C_{ij}(t) = \sum_{n=1}^{\infty} \Psi_i^{(n)} \Psi_j^{(n)*} e^{-m_n t} 
  \label{eq:spec_dec}
\end{eqnarray}
where $C(t)$ is the connected correlation matrix for distance $t$ in one of
the spatial directions.

On the lattice one can use only a truncated set of operators ${\cal{O}}_{i}
(i=1,\ldots,N)$ which is assumed to allow the extraction of the lowest lying
states by considering the eigenvalue problem for $C_{ij}(t)$.  Solving instead
the generalised eigenvalue problem
\begin{eqnarray}
\sum_j  C_{ij}(t) \Psi_j^{(n)} &=& \lambda^{(n)}(t,t_0) 
\sum_j C_{ij}(t_0) \Psi_j^{(n)}
  \label{eq:gen_eigen}
\end{eqnarray}
errors related to this truncation can be minimised ($t>t_0, t_0=0,1$)
\cite{luescher}.

The wavefunction of state $n$ in the given operator basis is found (at fixed
small distance $t>t_0$) to be
\begin{eqnarray} 
 \Psi_i^{(n)}(t)  = 
\langle \mathrm{vac} | ~{\cal {O}}_i ~e^{-H t} ~| {\bf \Psi}^{(n)} \rangle  \ .
\end{eqnarray} 
The components characterise the contribution of the original lattice operators
to the (ground or excited) state in the $J^{PC}$ channel.  Using operators of
different transverse extension (eventually only a subset of a used operator
set with fixed quantum numbers) one can derive their relative contribution to
an optimised wavefunction.  The masses of these states are obtained fitting
the diagonal elements $\mu^{(n)}(t)$
\begin{eqnarray} 
  \mu^{(n)}(t)= \sum_{ij} \Psi_i^{(n)*} C_{ij}(t)
  \Psi_j^{(n)} 
\end{eqnarray}
to $\cosh$ form.

Our base in the Higgs-channel ($0^{++}$) is built by the operators $\rho_x^2$,
$S_{x,\mu}(l)$ as well as quadratic Wilson loops of size $l \times l$.  In the
W-channel ($1^{--}$) we use the operators $V_{x,\mu}^b(l)$ and in the
($2^{++}$)-channel $S_{x,\mu}(l)-S_{x,\nu}(l)$.  Here the following notation
is used ($l$ counts the string length in lattice units)
\begin{eqnarray}
  S_{x,\mu}(l)=\frac{1}{2}{\mathrm{tr}}(\Phi^+_x U_{x,\mu}\ldots
  U_{x+(l-1)\mu,\mu}\Phi_{x+l\mu}),  \nonumber \\
  V_{x,\mu}^b(l)=\frac{1}{2}{\mathrm{tr}}(\tau^b\Phi^+_x U_{x,\mu}\ldots 
  U_{x+(l-1)\mu,\mu}\Phi_{x+l\mu}). \nonumber
\end{eqnarray}

\begin{figure}[!htb]
  \vspace{-7mm}
  \centering
  \epsfig{file=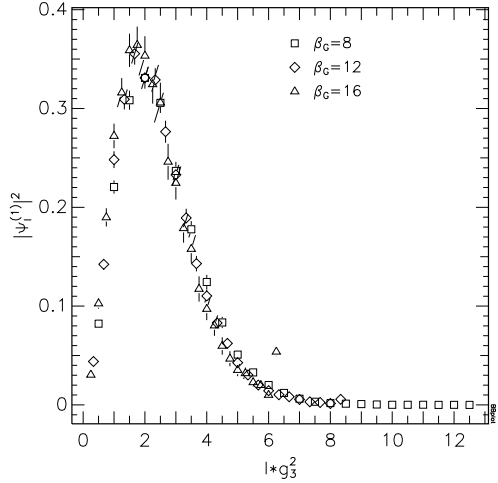,angle=0,width=65mm}
  \vspace{-1cm}
  \caption{\sl Squared  wavefunction $|\Psi_l^{(1)}|^2$ for $1^{--}$ groundstate 
    $n=1$ vs. physical length}
  \label{fig:ground}
  \vspace{-15mm}
\end{figure}

\begin{figure}[!htb]
  \centering
  \epsfig{file=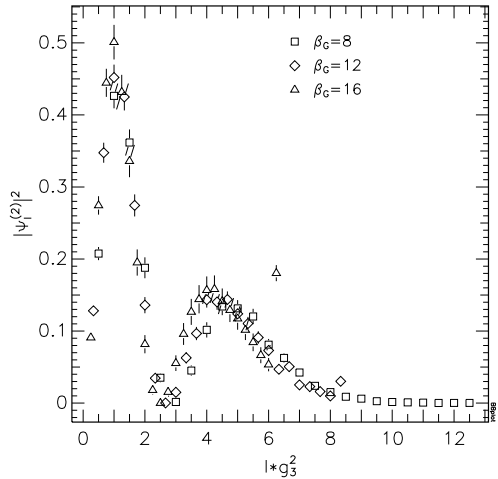,angle=0,width=65mm}
  \vspace{-1cm}
  \caption{\sl Same as  Fig.~3 for first excited state $n=2$}
  \label{fig:first}
  \vspace{-5mm}
\end{figure}

As an example we show the squared wavefunction for ground state and first
excitation in the $1^{--}$-channel in the high temperature phase near to the
transition as function of the operator length $l$ for a $50^3$ lattice. The
length is mapped to universal $g_3^2$ units in order to accommodate three
$\beta_G$ values at the line of constant physics $m_3^2/g_3^4 \approx 0.002$
(Figs.~\ref{fig:ground},\ref{fig:first}) 
at $M_H^*=70$~GeV.  A good scaling behaviour is observed.  The last point
($l=25$) seems to accumulate contributions from longer operators which do not
fit to the lattice.

\begin{figure}[!htb]
  \centering
  \epsfig{file=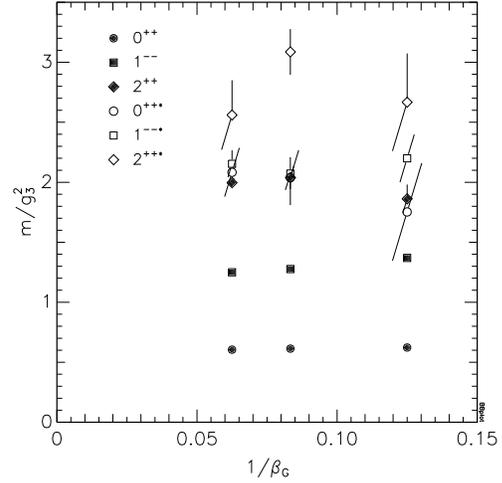,angle=0,width=65mm}
  \vspace{-1cm}
  \caption{\sl Masses of ground state and first excitation in physical units
    for different channels as function of $1/\beta_G\propto a$.}
  \label{fig:mass}
\end{figure}

First results on the spectrum using these wavefunctions are presented in Fig.
\ref{fig:mass} 
showing to what extent different masses are already approaching the continuum
limit.  Aspects of different smearing procedures, mixing properties of excited
states and results on different Higgs masses will be discussed in a future
publication \cite{future}.


\begin{thebibliography}{9}
\bibitem{wirNP97}
  M.~G\"urtler, E.M.~Ilgenfritz, J.~Kripfganz, H.~Perlt, and A.~Schiller,
  Nucl. Phys. B483 (1997) 383
\bibitem{generic} 
  K.~Kajantie, M.~Laine, K.~Rummukainen, and M.~Shaposhnikov,
  Nucl. Phys. B458 (1996) 90
\bibitem{muenster}  
  G.~M\"unster, 
  Nucl. Phys. B340 (1990) 559
\bibitem{kajantie} 
  K.~Kajantie, M.~Laine, K.~Rummukainen, and M.~Shaposhnikov,
  Nucl. Phys. B466 (1996) 189
\bibitem{surf} 
  M.~G\"urtler, E.M.~Ilgenfritz, and A.~Schiller, 
  hep-lat/9702020, to appear in Z.Phys.C
\bibitem{heitger}
  J.~Heitger, these proceedings
\bibitem{Philipsen} O.~Philipsen, M.~Teper, and H.~Wittig,
  Nucl. Phys. B469 (1996) 445
\bibitem{luescher}
  M.~L\"uscher and U.~Wolff, 
  Nucl. Phys. B339 (1990) 222
\bibitem{future}
  M.~G\"urtler, E.M.~Ilgenfritz, A.~Schiller, and C.~Strecha,
  in preparation 
\end{thebibliography}
\end{document}